\documentclass[%
 aps,
 amsmath,amssymb,
 reprint,%
]{revtex4-2}

\usepackage{graphicx}
\usepackage{subfig}
\usepackage{dcolumn}
\usepackage{bm}
\usepackage{amsmath}
\usepackage[dvipsnames]{xcolor}
\usepackage{orcidlink}%

\begin{document}

\preprint{PRD}

\title[Stochastic Mechanics of Hawking Radiation]{The Stochastic Mechanics of Hawking Radiation}
\thanks{This is a succeeding report to \textit{The Statistical Mechanics of Hawking Radiation}\cite{MacKay:2025yja}
}

\author{Noah M. MacKay \,\orcidlink{0000-0001-6625-2321}}
 \email{noah.mackay@uni-potsdam.de}
\affiliation{%
Institut für Physik und Astronomie, Universität Potsdam\\
Karl-Liebknecht-Straße 24/25, 14476 Potsdam, Germany
}%

\date{\today}

\begin{abstract}
In Ref. \cite{MacKay:2025yja}, Hawking radiation was analyzed through a statistical mechanics framework, revealing a structured microstate description of black hole horizons and information transfer into the radiation background. This study extends that approach by formulating Hawking radiation and black hole evaporation in the language of stochastic mechanics, employing an analytical Langevin framework and a numerical Euler iteration scheme. Both methods confirm that small black holes behave as thermal systems with Gaussian noise, while larger black holes develop a structured noise spectrum that aligns with the gradual contraction of the horizon. This suggests an alternative interpretation of Hawking radiation as an effective surface fuzziness, encoding horizon-scale fluctuations. The appendix provides a \textit{Wolfram Mathematica} blueprint for numerical simulations, open to heuristic modifications for further exploration of black hole noise spectra. 
\end{abstract}

\maketitle

\section{Introduction}

If a spherical body has a radius $r<r_S$, where  $r_S=2GM$ ($c=1$) is the Schwarzschild radius \cite{Schwarzschild:1916uq, Schwarzschild:1916ae}, such a body cannot exist in equilibrium and must undergo gravitational collapse into a central singularity enclosed by an event horizon. The radius of this region, the proper distance from the singularity to the horizon, is $r_S$ \cite{Penrose:1964wq}. This unstable, out-of-equilibrium entity is a black hole, and stationary black holes are predicted to emit Hawking radiation, gradually losing mass and eventually evaporating \cite{Hawking:1974rv, Hawking:1975vcx}.

Hawking radiation arises from the vacuum pair splitting of coupled (anti-)particles along the event horizon; one particle escapes as a relativistic particle with Compton mass $m_H=\hbar/(4\pi GM)$ \cite{MacKay:2025yja}, while the other falls into the black hole. This process establishes a thermodynamic system, where the black hole and radiation are often treated as being in thermal equilibrium with a shared temperature $T_H=\hbar/(8\pi GMk_B)$. It is an essential characteristic that the black hole temperature is inversely proportional to its mass. 

A heuristic statistical mechanical approach was introduced for black hole thermodynamics and Hawking radiation in Ref. \cite{MacKay:2025yja}. This species-independent formulation describes the emitted Hawking particle (and equivalently its ingoing trapped counterpart) through a worldline Lagrangian, suggesting a structured description of black hole microstates. In this framework, the evaporation rate of an individual microstate from the black hole surface to the radiation background is defined as Eq. (24) of Ref. \cite{MacKay:2025yja}, while the total number of microstates (given by Eq. (31) of Ref. \cite{MacKay:2025yja}) follows
\begin{equation} \label{number}
s=\frac{N}{8\pi^2}\ln\left(\frac{M}{m_P}\right),\quad N=\frac{A_\text{BH}}{16\pi l_P^2},
\end{equation}
where the Planck mass $m_P\equiv\sqrt{\hbar/G}=2.17\times10^{-8}$ kg and the Planck length $l_P\equiv\sqrt{\hbar G}=1.61\times10^{-35}$ m define fundamental quantum gravitational scales. The number of microstates scales with the quantum-gravitational area elements on the black hole horizon, implying that information storage is intrinsically linked to the surface degrees of freedom. This general approach applies to all statistical distributions -- as long as one Hawking particle occupies the emitted state --, while ensuring the conservation of microstates. Notably, this analysis recovers the conventional black hole temperature formula, reinforcing its thermodynamic consistency. 

One consequence of treating Hawking radiation as a statistical process is the ability to apply thermal noise analysis. In a conventional thermal gas, particles undergo Brownian motion due to frequent collisions with their nearest neighbors. However, since black hole temperature scales as $T\propto1/M$, (super-)massive black holes are extremely cold, suppressing significant thermal motion of emitted Hawking particles. Instead, their behavior is dominated by quantum fluctuations and vacuum effects, suggesting that zero-point energy fluctuations dictate noise properties in low-temperature regimes. 

Since black holes are inherently out-of-equilibrium systems, an open question is whether they exhibit self-organized criticality (SOC) -- a phenomenon where an unstable system self-regulates through sudden, chaotic avalanches \cite{Bak1987, Held1990}. Such avalanches often follow a $1/f$ noise spectrum, where the pwer spectrum density obeys a power law, $S(f)\propto f^{-\alpha}$, with $1<\alpha\leq2$. Here, $\alpha>1$ corresponds to L\'evy-Cauchy noise, and $\alpha=2$ corresponds to Gaussian noise. The presence of a power-law spectrum in black hole noise could indicate long-range correlations and horizon-scale fluctuations, making it a potential probe of dynamical black hole behavior.

Hawking radiation is conventionally described as thermal Gaussian noise ($\alpha=2$) \cite{Sonner:2012if}. However, deviations from Gaussianity may emerge due to fundamental black hole parameters. In particular, the spectral properties of emitted radiation could shift as the black hole evolves. While Hawking radiation is a gradual process, i.e. it does not directly trigger avalanches, it may act as a slow restabilization mechanism that drives the system toward equilibrium. Whether and how the noise spectrum transitions to avalanche-like behavior depends on underlying black hole dynamics, which this study will explore. 

Specifically, we construct the Langevin equation for Hawking radiation particles in Section \ref{sect:langeq}, and we discretize this equation into an Euler iteration scheme in Section \ref{sect:nums}. These sections provide the core analytical (Section \ref{sect:langeq}) and numerical (Section \ref{sect:nums}) results of this study. FInally, in Section \ref{sect:concl}, we conclude by discussing the implications of our findings for black hole noise and possible connections to SOC.

\section{The Langevin Equation} \label{sect:langeq}

As the black hole evaporates and its mass decreases, the corresponding rise in temperature could introduce stochastic effects in the emitted radiation. To examine whether such effects resemble thermal noise, we consider the standard Langevin equation \cite{lang}:
\begin{equation} \label{langeq}
\gamma\dot{x}(\tau)=-\nabla V(x)+ \xi(\tau),
\end{equation}
where $x=x(\tau)$ represents a stochastic trajectory in phase space rather than a physical length, and $\gamma$ is a damping coefficient. The first term on the right-hand side represents a restoring force from a potential well, $F(x)=-\nabla V(x)$, while the second term introduces a stochastic force, where $\xi(\tau)$ is a Gaussian noise term satisfying the expectation and correlation conditions $\langle\xi(t)\rangle=0$ and $\langle\xi_i(\tau)\xi_j(\tau') \rangle=2\gamma k_BT\Delta\nu\delta_{ij}(\tau-\tau')$. Here, $\Delta\nu$ is included to preserve the units of force-squared. Physically, it introduces a characteristic relaxation rate, preventing divergences in the noise spectrum and setting a characteristic relaxation time $\tau_\mathrm{eq}\equiv(\Delta\nu)^{-1}$. The noise amplitude squared is related to the the fluctuation-dissipation theorem as $\mathcal{A}^2\propto2\gamma k_BT$.

When a particle is ``kicked'' within $V(x)$ due to thermal fluctuations, the variance of the Gaussian noise term is related to the diffusion coefficient $D$, given by $D=k_BT/\gamma$ via the Stokes-Einstein-Sutherland relation, and the relaxation time $\tau_\mathrm{eq}$, such that $\sigma^2={2D\tau_\mathrm{eq}}$. This leads to a root-mean-square displacement of $x_{\mathrm{rms}}=\sqrt{2D\tau_\mathrm{eq}}$.
 
 In the context of Hawking radiation, a key question is how the black hole temperature (or equivalently, its size) affects the emitted radiation as a thermal noise spectrum. Does the stochasticity in the emitted radiation intensify for hotter (smaller) black holes while damping dominates for colder (larger) ones? Numerical analysis involves discretization and an Euler scheme to simulate the spectrum iteratively, while analytical analysis seeks to deduce the behavior of the noise spectrum by addressing the steady-state probability density. The expected form of this function depends on key parameters found in the Langevin equation and the physical constraints of the system.
 
 \subsection{Potential Well of a Black Hole}
 
In a steady-state potential well, a particle with zero-point energy can transition between discrete energy states. These transitions are described in the energy basis using creation and annihilation operators:
\begin{equation}
\widehat{H}_\varepsilon\left|\psi \right\rangle =a^{\dagger }_ja_k\left|\psi \right\rangle, 
\end{equation}
where \textit{k} is an occupied state and \textit{j} is an unoccupied state. 

\begin{figure}[h!]
\centering
\includegraphics[width=0.45\textwidth]{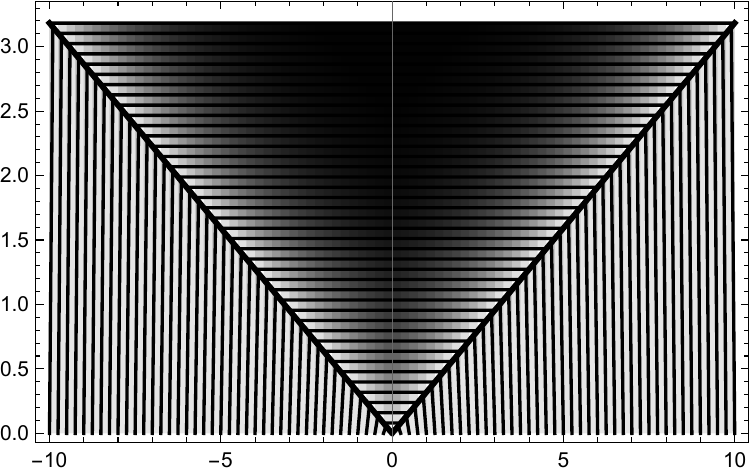}
\caption{\label{fig:horiz} The rectangular representation of an unraveled black hole. Thin, black rectangles of area $A=2\pi r_S^2$ are stacked with increasing $r_S$; the corners are connected by the V-shaped function $V(x)=|x|/\pi$ (bolded), with $|x|\in[0,r_{S,\text{max}}]$. Each horizontal line in the well (shaded region) represents a specific horizon, with the maximum horizon corresponding to the black hole's initial size. Evaporation follows a stepwise descent, transitioning to successively smaller horizons until reaching the corner node, where no horizon remains.}
\end{figure}

To model the statistical properties of Hawking radiation, we approximate the event horizon as an effective two-dimensional system. This follows from quantum statistical treatments of black holes, where the number of surface degrees of freedom scales with the black hole surface area. Specifically, we ``unravel'' the spherical event horizon onto an equivalent rectangular geometry: the circumference corresponds to the physical $x$-range $[-\pi r_S, \pi r_S]$, while the radial coordinate extends from $y=0$ at the geometric center to $y=r_S$ at the horizon. A visual aid is given as Figure \ref{fig:horiz}.

In this framework, intermediate states are assumed to be evenly spaced along $y$, with black hole evaporation corresponding to a gradual descent across these energy levels. Since the process is driven by a continuous loss of mass (equivalently, its size), we model the potential governing state transitions as a linear well:
\begin{equation}
\widetilde{V}(x)=F_0|x|,
\end{equation}
where $F_0$ is a constant restoring force driving state transitions. The emergence of this potential well will be justified in the subsequent Hamiltonian analysis.

Since this analysis is fully quantum mechanical, Maxwell-Boltzmann statistics are excluded. Instead, bosons and fermions obey the respective energy-basis (anti-)commutator relations:
\begin{eqnarray}
&\mathrm{Bose-Einstein:}\quad\left[a_m,a^{\dagger }_n\right]={\delta }_{mn}\hat{I},\nonumber\\
&\mathrm{Fermi-Dirac:}\quad\left\{a_m,a^{\dagger}_n\right\}=\delta_{mn}\hat{I},
\end{eqnarray}
where ${\delta }_{mn}$ is the Kronecker delta. These relations define the number operator ${\widehat{N}}_{mn}$, which governs state transitions and occupation. To model state transitions in the black hole system, we consider an initial reference state ($m=0$) evolving to an excited state $s$, governed by a sequence of ladder operations:
\begin{eqnarray}
\widehat{H}_{\varepsilon }\left|\psi \right\rangle& =a^{\dagger }_s\left[\left(a_{s-1}a^{\dagger }_{s-1}\right)\dots \left(a_1a^{\dagger }_1\right)\right]a_0\left|\psi \right\rangle\nonumber\\
& =\prod^{s-1}_{i=1}{\left(1+wn_i\right)}a^{\dagger }_sa_0\left|\psi \right\rangle, 
\end{eqnarray}
where $w=1$ for bosons and $w=-1$ for fermions. 

The discrete nature of black hole energy states follows naturally from the rectangular representation provided in Figure \ref{fig:horiz}, where each horizon corresponds to a step in a linear potential well. In this framework, ladder operators describe state transitions as an ascending mechanism from the bottom up, constructing the structured black hole well, while evaporation is the reversed descending mechanism from the maximum horizon downward.

The requirement $s\ge 2$, ensuring the validity of the product, implies that the black hole must possess at least two quantum states: a lower ``ground'' state at the geometric center and an upper ``excited'' state at the event horizon. Physically, this is to prevent having naked singularities for a structured black hole model, while only allowing emitted radiation states to be singular states. Using Eq. (\ref{number}), we find that for a black hole to support at least two states ($\lfloor s\rfloor=2$), it must have a minimum mass of $M=8m_P$. Below this threshold, the system no longer behaves as a structured potential well and instead transitions into an unbound radiation phase. These $8m_P$-scale black holes may represent a quantum gravitational regime distinct from the well-defined evaporation dynamics of larger black holes. The precise nature of this transition remains an open question.

While the current treatment distinguishes bosonic and fermionic treatments, this serves as a foundation for a species-independent description of black hole state transitions. A general formulation must account for both types of statistics while preserving the essential features of the underlying quantum system.

\subsubsection{Bosons}

If all Hawking particles are bosons ($w=1$), Bose-Einstein statistics allow multiple particles to occupy the same quantum state. At sufficiently low temperatures, bosons can condense into a single state rather than distributing across available states. If such an effect applies to the Hawking radiation of (super-)massive black holes, surface particles could preferentially occupy the lowest available states, such as along the horizon. This possibility suggests an enhancement mechanism for low-energy modes.

As a result, the single-particle Hamiltonian product expands into a sum:
\begin{equation}
\prod^{s-1}_{i=1}{\left(1+n_i\right)}=1+\sum_{j=1}^{s-1}n_j+\left(... \right)+\left\{n_1\left(... \right)n_{s-1}\right\}.
\end{equation}
Each term represents a different occupation configuration. However, due to the orthonormality condition $\langle i|j\rangle=\delta_{ij}$, contributions from terms involving products of multiple state populations vanish in the Hamiltonian. This leaves only the sum of individual state populations, corresponding to the total number of particles in the system when a single state is fully occupied. Thus, the system Hamiltonian $\mathcal{H}$ for bosonic Hawking particles is
\begin{equation}
\mathcal{H}_\mathrm{Boson}\left|\psi \right\rangle =\left(1+\sum^{s-1}_{i=1}{n_i}\right)a^{\dagger }_sa_0\left|\psi \right\rangle,
\end{equation}
where $\sum_i{n_i}=N$ represents the total number of particles and $(N+1)$ is the Bose enhancement factor.

To verify that a purely bosonic system has all of its particles in a single state, we must define the system's critical temperature. In statistical mechanics, thermal criticality marks a regime transition from a quantum statistical description ($T_\text{crit}>T$) to a semi-classical treatment under Maxwell-Boltzmann statistics ($T_\text{crit}<T$). 

Black hole evaporation in the Bose-Einstein description was previously studied, painting a quantum portrait of black holes as a leaky Bose condensate of gravitons \cite{Dvali:2011aa, Dvali:2012gb, Dvali:2012rt, Dvali:2014}. If we can define black hole criticality in this section as the point where additional quantum information deposited in the black hole is revealed in the radiation very rapidly \cite{Hayden:2007cs}, this critical point is reached when more than half of the system's internal entropy has evaporated. Via the Bekenstein-Hawking formula $S=k_B A_\text{BH}/(4l_P^2)$, this corresponds to the condition where the black hole's original area is at least halved:
\begin{equation}
A_\text{BH,crit}\leq\frac{A_\text{BH,0}}{2}\implies M_\text{crit}\leq\frac{M_0}{\sqrt{2}}.
\end{equation}
Since the Hawking temperature scales inversely with mass, the corresponding critical temperature is $T_\text{crit}\geq T_\text{0}\sqrt{2}$, which is greater than the system's initial temperature $T_0$. This suggests that if the emitted Hawking particles are purely bosonic, the surface particles they were initially paired with are in a condensate phase. However, the formation of a true black hole Bose condensate depends on whether interactions among the emitted particles allow for sufficient thermalization. The precise nature of such a condensate remains an open question, particularly in the context of off-equilibrium, non-isentropic Hawking radiation.

\subsubsection{Fermions}

If all Hawking particles are fermions ($w=-1$), their state occupancy is constrained by the Pauli exclusion principle. This prevents multiple fermions from occupying the same quantum state, ensuring non-trivial Hamiltonian dynamics. Extending this to a system of $N$ fermions, the single-particle Hamiltonian is summed over all particles:
\begin{equation}
{\mathcal{H}}_\mathrm{Fermion}\left|\psi \right\rangle =\sum_{l=1}^N\left[a^{\dagger }_sa_0\left|\psi \right\rangle\right]_l.
\end{equation}
Since each fermion must occupy a distinct state, the event horizon can be interpreted in two ways: either as a discretized set of surface states occupied by the total number of fermions or, heuristically, as permitting one fermion to exist in an intermediate state inside the black hole (following the visual aid in Figure \ref{fig:horiz}). This duality between surface state discretization and internal occupation is reminiscent of black hole mirroring effects and superfluid analogs \cite{Hayden:2007cs, Manikandan:2018urq}. 

A key indicator of this behavior is the system's Fermi energy,
\begin{equation}
\varepsilon_F=\frac{\hbar^2}{2m}\left(\frac{3\pi^2 N}{V_\text{BH}}\right)^{2/3},
\end{equation}
where $m$ is given by $m_H=\hbar/(4\pi GM)$, $N$ follows by Eq. (\ref{number}) and $V_\text{BH}$ is the effective black hole volume. The condition $\varepsilon_F>k_BT_H$ suggests that nearly all fermions in the black hole potential well occupy energy states below $\varepsilon_F$, forming a degenerate fermionic phase. Whether this configuration could support a fermionic condensate analogous to a Bardeen-Cooper-Schrieffer (BCS) superfluid remains an open question. If an effective attractive interaction exists -- perhaps mediated by gravitational effects --, it could lead to Cooper pairing and a superfluid-like behavior. However, the role of Hawking evaporation in maintaining or disrupting such a phase requires further investigation. \\

If Hawking particles, whether fermions or bosons, share an identical thermodynamic distribution in equilibrium -- such that their statistical effects are averaged out at asymptotically large $N$ --, then the fermionic and bosonic Hamiltonians approximate a convergence:
\begin{equation}
\mathcal{H}_\mathrm{Boson}\left|\psi \right\rangle\simeq{\mathcal{H}}_\mathrm{Fermion}\left|\psi \right\rangle =N\left(a^{\dagger }_sa_0\left|\psi \right\rangle\right).
\end{equation}
This convergence arises because, with $N\propto M^2$ via Eq. (\ref{number}), astrophysical black holes inherently possess large $N$, where thermodynamic averaging suppresses individual quantum effects. In this regime, statistical behavior dominates over species-dependent quantum properties, leading to an emergent classical description of collective black hole microstates. 

Since this formulation describes the entire black hole thermodynamic system, the characteristic energy scale is set by the Hawking temperature, $E=k_BT_H$, which serves an an approximate eigenvalue for both system Hamiltonians. This identification reflects an average energy per degree of freedom, reinforcing the quantum-classical correspondence in the large-$N$ limit. However, potential deviations may arise for smaller black holes, where finite-$N$ effects could preserve residual species-dependent behavior.

\subsubsection{Potential Well Profile}

As long as the reference state $0$ corresponds to the surface state $s$, the operator product $a^\dagger_i a_i$ represents the total energy of a single particle in physical space. Since the black hole's potential energy far exceeds the kinetic energy of a singularly emitted Hawking particle, the instantaneous potential energy at the event horizon is equivalent to the Hawking thermal energy eigenvalue: 
\begin{equation}
V(x_0)=\frac{\hbar}{8\pi GM}.
\end{equation}
This result is directly reproducible from the classical surface gravitational potential energy: $V(x)=GmM/x$, by substituting $m=m_H$ and $x=r_S$. This agreement reinforces the semiclassical consistency of the potential well approach and its connection to black hole thermodynamics.

To relate this to our effective 2D representation of an unraveled black hole, we note that as the event horizon contracts radially inward due to Hawking radiation, the black hole's surface undergoes a linear descent in state-space. If the potential gradient remains uniform, i.e., $-\nabla V(x)=\mathrm{constant}$, the system can be modeled by a linear potential well:
\begin{equation} \label{bhwell_vx}
V(x)=\frac{\hbar}{8\pi GM}\frac{|x|}{r_S}.
\end{equation}
Here, the scaling factor $|x|/r_S$ reflects the inward contraction of the horizon in the unraveled representation, ensuring a constant restoring force.

This formulation justifies the use of a linear potential well in the Langevin equation, as state transitions correspond to a steady, gradient-like inward motion along the radial dimension. The corresponding restoring force,
\begin{equation} 
F=\frac{\hbar}{8\pi GMr_S}=\frac{\hbar}{A_\text{BH}},
\end{equation}
establishes an inverse relationship with the black hole surface area. This scaling arises naturally because the restoring force per unit displacement is dictated by the system's geometry: a larger horizon area implies a weaker restoring effect, and consequently a longer time to achieve equilibirum. Since the characteristic relaxation time follows $\tau_\mathrm{eq}\propto A_{BH}/\hbar$, this aligns perfectly with the known timesclae for black hole evaporation \cite{Pickover:1996, Carroll:1996}. Because black hole evaporation is fundamentally governed by thermodynamic properties, the relaxation time inherits the same scaling: 
\begin{equation}
\tau_\mathrm{eq}\equiv\tau_\mathrm{evap}=320M\frac{A_{BH}}{\hbar},
\end{equation}
interpreting black hole evaporation as a long-timescale equilibration process within the system's state space.

\subsection{Obtaining Parameters}

\subsubsection{Diffusion and Damping}

In general, diffusion acts as the transport coefficient for mass flux, with the transport equation $J_i=-D\partial_i\varrho$, where $\varrho$ is the flowing mass density. Extending this to momentum transport, we introducing a flow velocity $u_j$ and define an off-diagonal momentum flux: $\pi_{ij}=-D\partial_i(\varrho u_j)$. Under the assumption of steady mass flow, such that $\pi_{ij}=-D\varrho\partial_i u_j$, this leads to a transport law for shear stress: $\pi_{ij}\equiv-\eta\partial_i u_j$. Here, $\eta$ is the shear viscosity, which relates to diffusion and mass density as
\begin{equation}
\eta=\varrho D.
\end{equation}
In the black hole system, the diffusion of Hawking radiation in the surrounding background can be mapped to the diffusion of the Hawking particles along the event horizon. Since mass density is thermodynamically equivalent to energy density $\epsilon$, and heat flow is linked to mass flow, it is natural to express mass density in terms of entropy: $\varrho=\mathcal{S} T_H$, where $\mathcal{S}=S/V$ is the entropy density. Using this, diffusion can be expressed in terms of the shear viscosity-to-entropy density ratio:
\begin{equation}
D=\frac{\eta}{\mathcal{S}T_H}.
\end{equation}
Applying the AdS/CFT correspondence for a black brane, the shear viscosity-to-entropy density ratio is known to be a universal constant \cite{Kovtun:2004de}: ${\eta}/\mathcal{S}={\hbar}/({4\pi k_B})$. Subsituting this into our diffusion expression, we find that the diffusion of Hawking radiation -- both in the surrounding radiation background and along the event horizon -- is inversely proportional to the black hole thermal energy:
\begin{equation}
D=\frac{\hbar}{4\pi k_BT_H}=2GM.
\end{equation}
Therefore, the corresponding damping coefficient $\gamma\equiv k_BT_H/D$ is directly proportional to the restoring force:
\begin{equation}
\gamma=\frac{\hbar}{16\pi G^2M^2}= F.
\end{equation}
This result implies that damping effects in the black hole system are naturally regulated by the same force responsible for state transitions. It also suggests a self-consistent thermodynamic balance, reinforcing the equivalence between relaxation processes and black hole evaporation dynamics.

\subsubsection{Variance and Noise Amplitude}

With the Gaussian noise variance given by $\sigma^2=2D\tau_\mathrm{eq}$, the corresponding variance for Hawking radiation, along with the root-mean-square displacement $\sigma$, takes the form
\begin{equation}
\sigma^2=\frac{80\,A^2_\text{BH}}{\pi\,l_P^2}\implies\sigma=\frac{4A_\text{BH}}{l_P}\sqrt{\frac{5}{\pi}},
\end{equation} 
This result demonstrates that the root-mean-square displacement scales with the Bekenstein-Hawking entropy $S=k_BA_\text{BH}/(4l_P^2)$, implying that fluctuations in Hawking radiation are fundamentally tied to the total number of accessible microstates. The associating noise amplitude, given by $\mathcal{A}=\sqrt{2\gamma k_BT_H/\tau_\text{eq}}$, yields
\begin{equation}
\mathcal{A}=\frac{F}{8\sqrt{20\pi}}\frac{m_P}{M},
\end{equation}
revealing that the noise amplitude scales with $\sim M^{-3}$, rather than just the temperature $T\propto1/M$ or the restoring force $F\propto 1/A_\text{BH}$ alone. This verifies the thermodynamic expectation that temperature governs the strength of fluctuations, but with an additional suppression factor for large black holes.

\subsection{Steady-State Probability Function}

For a standard Langevin equation describing $\alpha$-specific noise in a linear potential well, the corresponding steady-state Fokker-Planck equation yields a probablity distribution in the form of the complex Mittag-Leffler function \cite{MacKay:2024}. From a phenomenological perspective, where the noise amplitude $\mathcal{A}$ and restoring force $F$ satisfy the inequality relation dictated by the strength parameter $1<\alpha\leq2$ -- defining the stochasticity of a complex system --, the Mittag-Leffler distribution is given by:
\begin{equation} \label{alphapdf}
P_{st,\alpha}(x) =\frac{F}{2\mathcal{A}^\alpha} \frac{\Gamma(\alpha)^{1/(\alpha-1)}}{\Gamma\left(\alpha/(\alpha-1)\right)}\mathrm{E}_{\alpha-1}\left(-\frac{F}{\mathcal{A}^\alpha}|x|^{\alpha-1}\right).
\end{equation}
The behavior of this function depends on $\alpha$:
\begin{itemize}
\item For $\alpha=1$ (L\'evy noise), the probability distribution vanishes due to $F<\mathcal{A}$, indicating a regime where deviations from equilibrium dominate over restoration back to equilibirum. 

\item For $\alpha=2$, the distribution follows a double-sided exponential decay, corresponding to an overdamped Langevin process with $F>\mathcal{A}$.
\end{itemize}

 It should be noted that the dimensional consistency in the probability distribution is not explicitly verified in Ref. \cite{MacKay:2024}. This is particularly relevant for fractional values of $\alpha$, where the Mittag-Leffler function alters the effective dimensionality of the argument. A more detailed dimensional analysis may be necessary to ensure consistency in the physical interpretation of fractional stochastic regimes.

For the specific case of Hawking radiation as thermal noise, the obtained expressions for $F$ and $\mathcal{A}$ confirm that $F>\mathcal{A}$. This classifies the noise as Gaussian ($\alpha=2$), which aligns with the conventional understanding of Hawking radiation as Johnson-Nyquist thermal noise \cite{Sonner:2012if}. Given Eq. \ref{langeq} and ensuring dimensional consistency in the argument, the steady-state probability distribution takes the form 
\begin{equation}
P_{st}(x)=\frac{F}{2\gamma D} \exp\left(-\frac{F}{\gamma D}|x| \right).
\end{equation}
Applying this to the Hawking radiation scenario, where the restoring force $F=\hbar/A_\text{BH}$, the damping coefficient is $\gamma=F$, and the diffusion coefficient is $D=2GM$, the resulting probability distribution for Hawking radiation states is
\begin{equation}
P_{st}(x)=\frac{1}{2r_S} \exp\left(-\frac{1}{r_S}|x| \right),
\end{equation}
or alternatively $P_{st}(x)\propto\exp\left[-{V(x)}/({k_B T_H}) \right]$ provided Eq. (\ref{bhwell_vx}).

This result demonstrates that the probability distribution of state fluctuations is directly governed by the horizon scale, with a characteristic decay length on the order of the Schwarzschild radius $r_S$. This suggests that the steady-state properties of Hawking radiation fluctuations are inherently tied to the spatial structure of the black hole, reinforcing the increasingly localized nature of fluctuations as the black hole shrinks. However, this localization raises an important question: as $r_S$ decreases, do semi-classical approximations remain valid, or do quantum gravitational corrections become significant at sufficiently small scales? 

Moreover, this result confirms that the probability function approaches zero for large masses, aligning with thermodynamic intuition -- lower temperatures suppress thermal agitation. To analyze this distribution, we express the black hole radius in terms of integer multiples of the Planck length, $r_S=bl_P$. Substituting this into the probability density function yields
\begin{equation} \label{rspdf}
P_{st}(x)\mathrm{\left[\frac{1}{m}\right]}=\frac{10^{35}}{3.2b}\exp\left(-\frac{10^{35}}{1.6b}\frac{|x|}{\mathrm{m}} \right).
\end{equation}
This expression reveals two asymptotic behaviors:
\begin{itemize}
\item For small $b$ (small black holes), the prefactor $1/l_P$ dominates, leading to pronounced fluctuations.
\item For large $b$ (large black holes), the term $b$ dominates, suppressing fluctuations and reducing probability amplitudes. 
\end{itemize}

\begin{figure} [h!]
\centering
\includegraphics[width=0.45\textwidth]{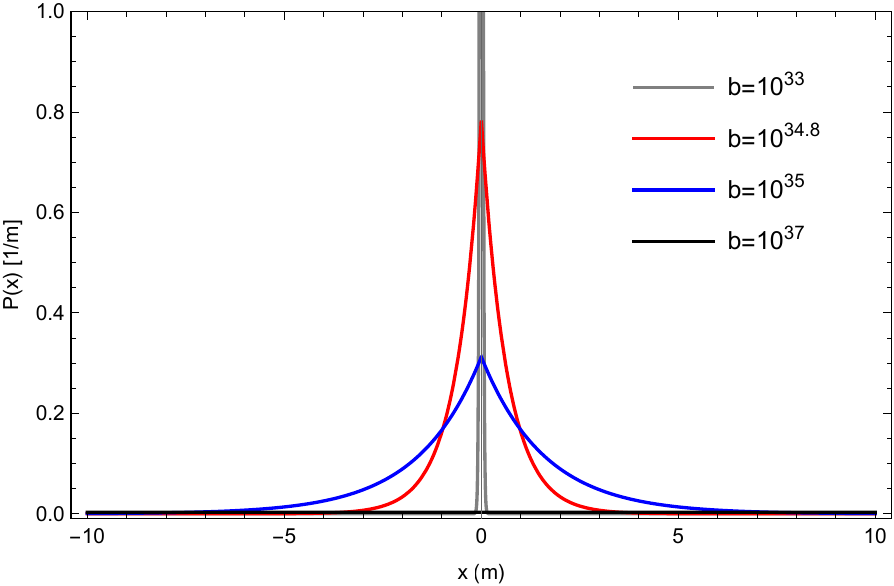}
\caption{ \label{probgraph} Hawking radiation steady-state probability distributions for sample radii $r_S= bl_P$: $b=10^{33}$ (gray), $b=10^{34.8}$ (red), $b=10^{35}$ (blue), $b=10^{37}$ (black).}
\end{figure}

To visualize these effects, we select sample values of $b$ that maintain the prefactor near $\sim10^{0}$ in order of magnitude. The resulting distributions are plotted in Figure \ref{probgraph}. Notably, for $b\geq10^{37}$ (corresponding to black hole radius $r_S\geq160\,\mathrm{m}$), the distribution flattens to near zero, characteristic of L\'evy-Cauchy noise. Conversely, for $b\leq10^{33}$ ($r_S\leq0.016\,\mathrm{m}$), the distribution exhibits a sharply-peaked Dirac delta-like form.

A key observation is that, while Hawking radiation follows Gaussian noise statistics, the black hole mass modulates the strength of fluctuations without changing the stability index $\alpha$. Consequently, this verifies that large black holes behave classically with negligible fluctuations, while imposing that smaller black holes are noise-dominated and that late-stage evaporation could become erratic via the $\mathcal{A}\propto M^{-3}$ scaling. These results suggest that black hole fluctuations become increasingly localized as the black hole shrinks, emphasizing the transition from stable semi-classical behavior to a fluctuation-dominated regime.

 If Gaussian noise persists down to the Planck scale, this could indicate that quantum gravity does not introduce new stochastic effects -- only stronger fluctuations. However, if deviations from Gaussianity emerge at small masses, this may signal a fundamental change in black hole physics near the Planck scale.
  
\section{Euler Scheme and Numerics} \label{sect:nums}

Similating the stochasticity of a particle governed by Eq. (\ref{langeq}) is possible via a forward Euler iterative scheme \cite{Yuvan2021, Yuvan2022ent, Yuvan2022sym, Bier2024}. For a linear potential well $V(x)=F|x|$, we follow the more general $\alpha$-dependent numerical procedure provided in Ref. \cite{MacKay:2024}, using $\alpha=2$ for the Gaussian noise case. Since stochastic agitation occurs through discrete kicks, we compute iterations over a sequence of ``timesteps," where each timestep corresponds to a random displacement at the $i$-th moment in time. This requires transforming the Langevin equation from a differential equation into an iterative update equation:
\begin{equation}
\Delta x_i=\frac{\Delta\tau}{\gamma}\left[-F\mathrm{sgn}(x_i)+\mathcal{A}\frac{\theta_{i}}{\sqrt{\Delta \tau}}\right],
\end{equation}
where $\Delta x_i=x_{i+1}-x_i$, $F(x_i)=-F\mathrm{sgn}(x_i)$ is the restoring force term for a given $x_i$, and $\theta_{i}$ is a random value drawn from a Gaussian distribution at each timestep. Rearranging, we define our Euler scheme by isolating $x_{i+1}$ on the left-hand side.

Because the steepness of the V-shaped linear well depends on $F$, the Euler scheme can become inaccurate for large $F$, where large restoration forces dampen the fluctuations due to an exceedingly large slope. This issue can be mitigated by reducing $\Delta \tau$, ensuring that $\Delta x_i=x_{i+1}-x_i$ is just as small. This effectively treats the discrete iterations as a continuous trajectory. However, this comes at the cost of increased computational complexity: a continuous noise spectrum requires a lerge number of timesteps to accurately capture stochastic fluctuations, e.g. $(\Delta\tau)^{-1}$ quantitatively. Balancing numerical stability with computational efficiency remains a key challenge, and adaptive timestep strategies may provide a way to optimize simulation performance.

Given the expressions for $\gamma$ and $\mathcal{A}$ in terms of $F$, and defining a mass scale $M=nm_P$, our Euler scheme has an effectively normalized restoring force:
\begin{equation} \label{it1}
x_{i+1}=x_i+\Delta\tau \left[-\mathrm{sgn}(x_i)+\frac{1}{8\sqrt{20\pi}n}\frac{\theta_{i}}{\sqrt{\Delta \tau}}\right].
\end{equation}
One can see that for large $n$ (large black hole mass), the noise amplitude decreases, maintaining consistency with thermodynamic convention and aligning with the logic of Eq. (\ref{rspdf}). 

To examine the noise spectrum for an entire system of Hawking particles, we factor in the total number of emitted particles, $N\equiv M^2/m_P^2= n^2$, while keeping the normalized restoring force unitary due to its relationship with the relaxation time. Defining the square-bracket expression in Eq. (\ref{it1}) times $\Delta \tau$ as $\text{dx}_i$, we note that that for large $n$, the noise term dominates over the restoring force. This produces a L\'evy-Cauchy, non-equilibirum noise characteristic, which also aligns with Eq. (\ref{rspdf}) in such a case. As a result, the ``L\'evy-ness'' of the whole-system noise spectrum is maintained by a $1/n$ scaling on $\text{dx}_i$. The Euler scheme then takes the form:  
\begin{equation} \label{it2}
x_{i+1}=x_i+\frac{\Delta\tau}{n} \left[-\mathrm{sgn}(x_i)+\frac{n}{8\sqrt{20\pi}}\frac{\theta_{i}}{\sqrt{\Delta \tau}}\right].
\end{equation}

To implement this Euler scheme in \textit{Mathematica}, we use the commands \texttt{RandomVariate} and \texttt{NormalDistribution}, along with a few lines of code to generate a large set of Gaussian-distributed numbers, thereby simulating Eq. (\ref{langeq}). The \texttt{RandomVariate} command requires a set range for the Gaussian distribution from which random values are drawn, meaning the range must match the number of timesteps as the \texttt{length} of the noise realization. Allowing $\mathtt{length}\propto(\Delta\tau)^{-1}$ quantitatively, a small $\Delta \tau$ thus corresponds to a large number of timesteps to ensure sufficient simulation of the noise spectrum.

For Hawking radiation, the black hole mass scale inherently determines the noise simulation length, as a smaller mass has a shorter lifespan and a larger mass has a longer lifespan. To incorporate this, we heuristically set $\mathtt{length=10^3*Ceiling[Log[n]]}$, where the factor $\mathtt{10^3}$ ensures a sufficiently long simulation, and the \texttt{Ceiling} function prevents the systematic error of fractional ranges in \texttt{RandomVariate}. This choise ensures that the number of iterations grows logarithmically with mass, maintaining computational feasibility for very large $n$. By construction, $\Delta\tau=1/\mathtt{length}$, which introduces a smallness problem in $\textrm{dx}_i$, effectively yielding $x_{i+1}\simeq x_i$. To perserve noise variations despite small $\Delta\tau$, we scale $\textrm{dx}_i$ by the heuristic factor of $\mathtt{10^6}$, ensuring that stochastic deviations remain significant while keeping $\Delta\tau\propto\mathtt{10^{-3}}$ for numerical stability. Future optimization may involve adjusting this scaling dynamically based on noise spectrum characteristics. 

\begin{figure} [h!]
\centering
\includegraphics[width=0.45\textwidth]{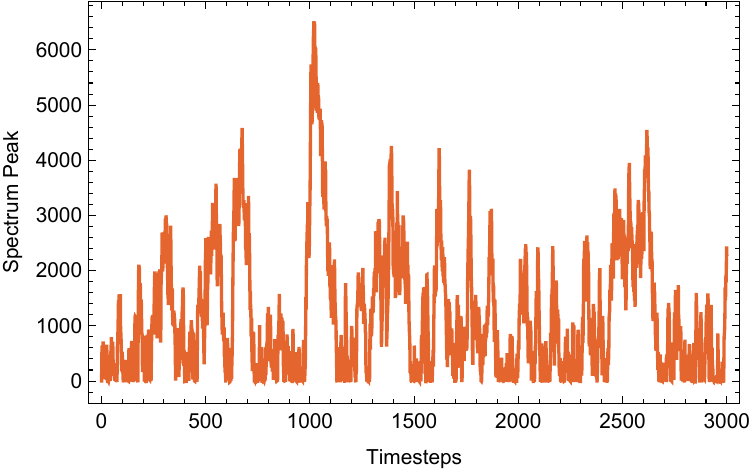}
\caption{ \label{fig:its1} Noise spectrum for mass scale $n=8$ ($M=8m_P$). With an associating temperature of $5.66\times10^{30}$K, the noise is strongly thermal, exhibiting an uncorrelated Gaussian profile.}
\end{figure}

\begin{figure} [h!]
\centering
\includegraphics[width=0.45\textwidth]{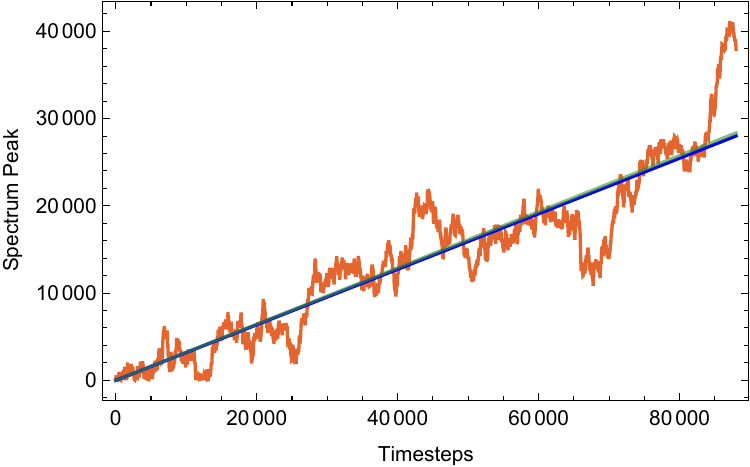}
\caption{ \label{fig:its2}  Noise spectrum for mass scale $n=10^{38}$ ($M\simeq2\times10^{30}\equiv M_\odot$). With an associating temperature of $0.617\,\mathrm{nK}$, the noise  remains fluctating but is non-thermal. The blue solid line represents $y=x/\pi$, while the green transparent line is the linear fit of the noise, $y_\text{fit}=0.3223 x$, showing a 0.988 1:1 ratio with the reference.}
\end{figure}

\begin{figure} [h!]
\centering
\includegraphics[width=0.45\textwidth]{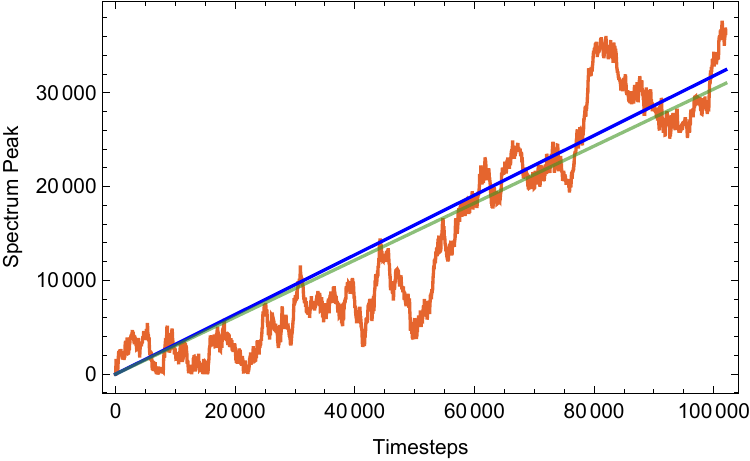}
\caption{ \label{fig:its3} Noise spectrum for mass scale $n=10^{44}$ ($M\simeq2\times10^{36}\equiv 10^6M_\odot$). With an associating temperature of $6.17\times10^{-15}\,\mathrm{K}$, the noise spectrum is undeniable non-thermal and quantum mechanical. The blue solid line represents $y=x/\pi$, while the green transparent line is the linear fit of the noise, $y_\text{fit}=0.3042 x$, yielding a 0.956 1:1 ratio.}
\end{figure}

The profile and behavior of the noise simulation depend on the mass scale $n$. As a proof of concept, we expect that small masses (hot temperatures) produce a noise spectrum akin to white (Gaussian) noise. However, the behavior of larger masses (colder temperatures) is less intuitive. Since the black hole is assigned a linear well profile, the generated noise may exhibit an ``uphill climb'' effect, reflecting the amplified deviations in $\mathrm{dx}_i$. Figures \ref{fig:its1}--\ref{fig:its3} show the noise spectra under various mass scalings: $n=8$ for Figure \ref{fig:its1}, $n=10^{38}$ for Figure \ref{fig:its2}, and $n=10^{44}$ for Figure \ref{fig:its3}.

 For $n=8$ in Figure \ref{fig:its1}, the noise is characteristically thermal, preserving time symmetry and upholding Onsager's principle of microscopic reversibility \cite{Onsager1931pr37, Onsager1931pr38}. This behavior is consistent with small black holes having purely thermal Hawking radiation noise, where fluctuations are uncorrelated and Gaussian.
 
  For larger black holes, such as $n=10^{38}$ and $n=10^{44}$ in Figures \ref{fig:its2} and \ref{fig:its3}, respectively, the noise spectra exhibit a linear climb while maintaing stochastic deviations from the linear trend. This trend follows the function $y=x/\pi$, which resembles the positive half of the V-shaped function $V=|x|/\pi$ shown in Figure \ref{fig:horiz}. Given this, the non-thermal noise spectra for large masses reflect the structure of the black hole's event horizon itself. The extreme deviations from the static line correspond to anomolies inherent in the stochastic noise generation, where fluctuations encode random perturbations to the black hole's surface.

It is important to emphasize that this simulation serves as a proof-of-concept for black hole evaporation and the stochastic nature of noise at the event horizon. The fluctuations around the linear trend suggest a ``fuzziness'' in the black hole surface, which aligns with the string-theoretical fuzzball model of black holes \cite{Lunin:2001jy, Lunin:2002qf, Mathur:2002ie, Mathur:2005zp, Mathur:2008wi, Mathur:2008nj, Mathur:2009hf, Mathur:2012zp}. Extreme deviations from the linear trajectory could indicate anisotropies in the black hole system:
\begin{itemize}
\item Glitches in axial rotations \cite{Hujeirat:2024a, Hujeirat:2024b}: Fluctuations below the linear trend may correspond to Kerr-like contraction effects, where stochastic deviations alter the angular momentum distribution of a spinning black hole.
\item Mass intake effects: Fluctuations above the linear trend may correspond to small accretion effects, where stochastic changes in mass induce perturbations in the horizon geometry, consistent with $dr_S\propto dM$.
\end{itemize}

These findings suggest that black hole noise is not purely Gaussian across all mass scales -- rather, horizon-scale fluctuations introduce a structure that encodes geometry and dynamical properties of the black hole. Future work may explore whether these deviations serve as a signature of horizon-scale quantum effects or a precursor to stochastic deviations in black hole thermodynamics.

\section{Conclusion} \label{sect:concl}

In this study, we explored the stochastic properties of Hawking radiation using a Langevin framework, revealing that small black holes exhibit Gaussian white noise, while large black holes develop structured noise spectra. Notably, the linear trend in large-mass noise aligns with the V-shaped horizon potential, suggesting that black hole noise retains an imprint of its geometric structure. Extreme fluctuations may correspond to horizon-scale anisotropies, such as spin variations or mass intake effects.

Thermodynamically, our results confirm that fluctuation amplitude scales inversely with mass, but in the large-mass regime, deviations from Gaussianity emerge, hinting at a possible transition to self-organized criticality (SOC). This raises open questions about long-range correlations in Hawking emission and potential observational signatures of quantum gravitational effects.

Future work could explore power spectral density analyses to test for SOC-like behavior and use numerical relativity to investigate how stochastic noise encodes horizon-scale dynamics. These findings suggest that black hole noise is not merely thermal, but may serve as a proble into black hole structure, information dynamics, and quantum gravity effects.


\appendix

\section{Layout of the Euler Iteration Scheme}

This appendix provides a blueprint to implementing the Euler iteration scheme in \textit{Wolfram Mathematica} to generate black hole noise simulations. The goal is to translate the numerical approach from Section \ref{sect:nums} into a structured computational procedure, incorporating the necessary heuristic choices.

We begin by defining the essential input parameters:
\begin{verbatim}
length = 10^3*Ceiling[Log[n]];
dt = 1/length;
data = RandomVariate[NormalDistribution[0, 1], 
	length];
kicks = Table[(data[[i]]/(dt^(1/2))),
	{i, length}];

AbsForce = 1; 
Amplitude = n/(8*Sqrt[20 Pi]);
Force[u_] := If[u > 0, -AbsForce, AbsForce]

n = ## (*insert value*);
\end{verbatim}
Here, $\mathtt{n}=M/m_P$ represents the black hole mass scale. If mass contraints are imposed, an upper limit could be the mass of the observable universe, approximately $10^{54}$ kg.

Now we implement the iteration scheme using a loop structure with \texttt{Do} commands and imbedded \texttt{If} statements:
\begin{verbatim}
x[1] = 0.0000001;

Do[
	dx = (Force[x[i]] + Amplitude*kicks[[i]]) 
		dt*(10^6/b);
 	If[x[i] == x[1] && 
 		Amplitude*kicks[[i]] <= Force[x[i]], 
 		dx = 0];
 	x[i+1] = x[i] + dx;
 	If[x[i+1] < 0, x[i+1] = 0],
 		{i, 1, length}
 ]
 	
PosTable = Table[x[j], {j, length}];
f = Interpolation[PosTable];

Plot[{f[y]}, {y, 1, length}, Frame -> True, 
 PlotStyle -> {Hue[0.05, 0.8, 0.9]},
 FrameStyle -> Directive[Black], 
 PlotRange -> {All}, 
 FrameLabel -> {
 	{"Spectrum Peak", None}, 
 	{"Timesteps", None}
 }]
\end{verbatim}

The \texttt{Do} loop iterates over timesteps, applying the Euler update rule to evolve the position. The \texttt{If} conditions filter out invalid noise values that do not meet the simulation constraints. The generated positions are stored in \texttt{PosTable}, which can be plotted directly using \texttt{ListLinePlot} or interpolated into a smooth function for visualization. It is recommended to compile the loop at least twice to prime the code and produce results.

Since the Langevin framework relies on the linear well representation of black hole evaporation, the expected trend in the noise spectrum follows the  linear function $y=x/\pi$, especially for large black holes (large \texttt{n}). To validate this, we compare the simulation results with both the expected trend and a fitted linear model:
\begin{verbatim}
g = Fit[PosTable, {y}, y];

Plot[{f[y], y/Pi, g}, {y, 1, length}, 
 Frame -> True, 
 PlotStyle -> {Hue[0.05, 0.8, 0.9], 
   Blue, 
   {Hue[0.29, 0.77, 0.58], Opacity[0.6]}}, 
 FrameStyle -> Directive[Black], 
 PlotRange -> {All}, 
 FrameLabel -> {
 	{"Spectrum Peak", None},
 	{"Timesteps", None}
 }]
\end{verbatim}

Due to the stochastic nature of the noise simulation, exact reproduction of the figures in this study is not expected. However, the overall structure of the results should consistently exhibit a linear trend for large \texttt{n} and thermal Gaussian noise for small \texttt{n}. This linear trend should follow the expected function $y=x/\pi$, with the fitted line plotted to compare with the theoretical predictions.

\end{document}